\documentstyle[twoside,fleqn,espcrc2,epsf]{article}
\title{Instanton, Monopole Condensation and Confinement}
\author{H.~Suganuma
\address{Research Center for Nuclear Physics (RCNP), 
Osaka University, Ibaraki 567, Japan},
 M.~Fukushima $^{\rm a}$, H.~Ichie $^{\rm a}$ and A.~Tanaka $^{\rm a}$}

\begin{document}

\begin{abstract}
The confinement mechanism in the nonperturbative QCD is studied in 
terms of topological excitation as QCD-monopoles and instantons. 
In the 't~Hooft abelian gauge, QCD is reduced into an abelian 
gauge theory with monopoles, and the QCD vacuum can be regarded 
as the dual superconductor with monopole condensation, which 
leads to the dual Higgs mechanism. 
The monopole-current theory extracted from QCD is found to have 
essential features of confinement. 
We find also close relation between monopoles and instantons 
using the lattice QCD. 
In this framework, the lowest $0^{++}$ glueball (1.5 $\sim$ 1.7GeV) 
can be identified as the QCD-monopole or the dual Higgs particle. 
\end{abstract}

\maketitle

\section{\bf Dual Higgs Theory for NP-QCD}

Quantum chromodynamics (QCD) is established as the strong-interaction 
sector in the Standard Model, and the perturbative QCD 
provides the powerful and systematic method in analyzing high-energy 
experimental data. 
However, QCD is a `black box' in the infrared region still now 
owing to the strong-coupling nature, 
although there appear rich phenomena as 
{\it color confinement, dynamical chiral-symmetry breaking} and 
{\it topological excitation} in the nonperturbative QCD (NP-QCD).
In particular, 
confinement is the most outstanding feature 
in NP-QCD, and to understand the confinement 
mechanism is a central issue in hadron physics.

In 1974, Nambu [1] presented an interesting idea that 
quark confinement and string picture for hadrons 
can be interpreted as the squeezing of the color-electric flux 
by the {\it dual Meissner effect}, which is similar to formation 
of the Abrikosov vortex in the type-II superconductor.
This {\it dual superconductor picture} for the NP-QCD vacuum 
is based on the {\it duality} in the Maxwell equation, and 
needs condensation of {\it color-magnetic monopoles}, which is 
the dual version of electric-charge (Cooper-pair) condensation 
in the superconductivity. 

In 1981, 't~Hooft [2] pointed out that color-magnetic monopoles 
appear in QCD as topological excitation in the {\it abelian gauge} 
[2,3], which diagonalizes a gauge-dependent variable $X(s)$.
Here, SU($N_c$) gauge degrees of freedom is partially fixed 
except for the maximal torus subgroup U(1)$^{N_c-1}$ 
and the Weyl group. 
In the abelian gauge, QCD is reduced into 
a U(1)$^{N_c-1}$-gauge theory, 
and {\it monopoles with unit magnetic charge} appear at 
hedgehog-like configurations 
according to the nontrivial homotopy group, 
$\Pi _{2}\{{\rm SU}(N_c)/{\rm U}(1)^{N_c-1}\}$
=$Z^{N_c-1}_\infty$ [4-6].

In 90's, the Monte Carlo simulation based on the lattice QCD 
becomes a powerful tool for the analysis of the confinement mechanism 
using the maximally abelian (MA) gauge [7-15], which is 
a special abelian gauge minimizing the off-diagonal 
components of the gluon field.
Recent lattice studies with MA gauge have indicated 
monopole condensation in the NP-QCD vacuum [7-9] 
and the relevant role of abelian degrees of freedom, 
{\it abelian dominance} [9-12], for NP-QCD. 
In the lattice QCD in MA gauge, {\it monopole dominance} for NP-QCD is 
also observed as the essential role of QCD-monopoles for the linear 
quark potential [10], chiral symmetry breaking [11,12] 
and instantons [6,13,14].

In this paper, we study QCD-monopoles in the NP-QCD 
vacuum using the SU(2) lattice QCD in MA gauge. 
Next, we study the role of QCD-monopoles to quark confinement 
using the monopole-current theory extracted from QCD [16]. 
Finally, we study the correlation between instantons and QCD-monopoles 
in terms of remaining nonabelian nature in MA gauge.

\section{\bf Lattice QCD in MA Gauge}

The maximally abelian (MA) gauge is the best 
abelian gauge for the dual superconductor picture for NP-QCD. 
In this section, we consider the mathematical structure of MA gauge. 
In the SU(2) lattice formalism, MA gauge is defined 
so as to maximize 
\begin{eqnarray}
R\! &\equiv &\! \! \sum_{s,\mu}
  {\rm tr}\{U_\mu(s)\tau_3 U^{-1}_\mu(s) \tau_3\} \nonumber \\
 \! &= &\! \!2 \sum_{s,\mu}
  \{U_\mu^0(s)^2+U_\mu^3(s)^2-U_\mu^1(s)^2-U_\mu^2(s)^2\} \nonumber \\
\!
\end{eqnarray}
by the gauge transformation.
Here, $U_\mu(s)\equiv \exp\{ia eA_\mu(s)\} \equiv 
U_\mu^0(s)+i \tau^a U_\mu^a(s)$ 
denotes the link-variable on the lattice with spacing $a$. 
In MA gauge in the lattice formalism,
\begin{eqnarray}
X[U_\mu(s)] 
\equiv \sum_{\mu=1}^4 U_{\pm \mu}(s \! \! \! \! \! \! & ) & \! \! \! \! \! \! 
\tau _3U^{-1}_{\pm \mu}(s) 
\nonumber \\
= \sum_{\mu=1}^4
\{U_\mu(s)\tau _3U^{-1}_\mu(s) \! \! \! \! \! & + &\! \! \! \! \!
U^{-1}_\mu(s-\hat \mu )\tau _3U_\mu(s-\hat \mu )\}
\nonumber \\
\!
\end{eqnarray}
is diagonalized. 
Here, we use the convenient notation as 
$U_{-\mu}(s)\equiv U^{-1}_\mu(s-\hat \mu)$.

In MA gauge, there remain U(1)$_3$-gauge symmetry and 
global Weyl symmetry [13], because 
$R$ is invariant under the gauge transformation 
$
U_\mu(s) \rightarrow  v(s)U_\mu(s)v^{-1}(s+\hat \mu) 
$
with $v(s)=$ 
$e^{i\tau^3\phi^3(s)} \in {\rm U}(1)_3$ 
and the Weyl transformation 
$
U_\mu(s)\rightarrow WU_\mu(s)W^{-1} 
$
with $W \in {\rm Weyl}_2 \simeq Z_2$ being $s$-independent. 
$W$ is expressed as 
\begin{eqnarray}
W \! \! &\equiv & \! \! 
\exp\{i\pi({\tau_1 \over 2} \cos\phi +{\tau_2 \over 2} \sin\phi )\} 
\nonumber \\
  \! \! &= & \! \!
i(\tau_1 \cos\phi +\tau_2 \sin\phi )
=i \pmatrix {
0                &    e^{-i\phi }   \cr 
e^{i\phi }          &    0              
}\! \! \! , 
\end{eqnarray}
and interchanges SU(2)-quark color, 
$|+ \rangle=(^1_0)$ and $|- \rangle=(^0_1)$.
In the SU($N_c$) case, this Weyl symmetry 
Weyl$_{N_c}$ corresponds to the permutation group 
$P_{N_c}=Z_{N_c(N_c-1)/2}$, 
whose element interchanges 
SU($N_c$)-quark color [13,17].

Nonabelian gauge symmetry 
$G$ $\equiv {\rm SU}(N_c)_{\rm local}$ is reduced into 
$H \equiv {\rm U(1)}^{N_c-1}_{\rm local} 
\times {\rm Weyl}_{N_c}^{\rm global}$ in MA gauge. 
Then, the independent set of the gauge function 
$\Omega _{\rm MA}(s)$ which realizes MA gauge fixing 
corresponds to the coset space $G/H$: 
$\Omega _{\rm MA}(s) \in G/H$. 
The representative element 
for the link-variable in MA gauge is expressed 
as $U_\mu^{\rm MA}(s)\equiv 
\Omega _{\rm MA}(s) U_\mu(s)\Omega _{\rm MA}^{-1}(s+\hat \mu)$, 
and also forms $G/H$. 
Thus, MA gauge fixing obeys the nonlinear representation on 
coset space $G/H$.

The MA gauge function $\Omega _{\rm MA}(s) \in G/H$ is transformed 
nonlinearly by $g(s) \in G \equiv {\rm SU}(N_c)$ 
as $\Omega _{\rm MA}(s)\rightarrow \Omega _{\rm MA}^g(s)=h[g](s)\Omega _{\rm MA}(s)g^{-1}(s)$, 
where $h[g](s) \in H$ appears so as to satisfy 
$\Omega _{\rm MA}^g(s) \in G/H$. 
Actually, the successive gauge transformation, 
$\Omega _{\rm MA}^g$ after $g$, is equivalent to $h[g]\Omega _{\rm MA}$, 
and maximizes $R$.

According to the nonlinear transformation in $\Omega _{\rm MA} \in G/H$, 
any operator $\hat O_{\rm MA}$ defined in MA gauge 
transforms nonlinearly 
as $\hat O_{\rm MA}\rightarrow \hat O_{\rm MA}^{h[g]}$ 
by the SU($N_c$)-gauge transformation $g \in G$. \\
\noindent
(Proof) For simplicity, original $\hat O$ is assumed 
to obey the adjoint transformation by the SU($N_c$)-gauge 
transformation. 
Then, one finds $\hat O_{\rm MA} = \Omega _{\rm MA} \hat O \Omega _{\rm MA}^{-1}$. 
By $g \in G$, 
$\hat O_{\rm MA}$ is transformed as 
$\hat O_{\rm MA} \rightarrow  \hat O_{\rm MA}^g = 
\Omega _{\rm MA}^g \hat O^g {\Omega _{\rm MA}^g}^{-1}
=h[g]\Omega _{\rm MA}g^{-1} \cdot
g \hat O g^{-1} \cdot
g \Omega _{\rm MA} h[g]^{-1}
= h[g] \hat O_{\rm MA}h[g]^{-1} 
=\hat O_{\rm MA}^{h[g]}$ with $h[g] \in H$. 
This proof can be generalized 
to any operator $\hat O$.

If $\hat O_{\rm MA}$ is $H$-invariant, one gets 
$\hat O_{\rm MA}^{h[g]}=\hat O_{\rm MA}$ for any $h[g]\in H$, 
so that $\hat O_{\rm MA}$ is invariant under arbitrary 
gauge transformation by $g \in G$. 
Thus, one finds a useful criterion on the SU($N_c$)-gauge invariance 
of the operator in MA gauge [13].\\
\noindent
``If an operator $\hat O_{\rm MA}$ defined in MA gauge is 
$H$-invariant, $\hat O_{\rm MA}$ is proved to be also invariant 
under the whole gauge transformation of $G$.''

\section{\bf Abelian/Monopole Projection}

The SU(2) link-variable $U_\mu(s)$ can be factorized as 
$U_\mu(s)=M_\mu(s)u_\mu(s)$, where 
$u_\mu (s) \equiv \exp\{i\tau^3\theta ^3_\mu (s)\} \in {\rm U}(1)_3$ 
is {\it abelian link-variable} and 
$M_\mu (s)\equiv e^{i\tau ^1\theta _\mu ^1(s)+\tau ^2\theta _\mu ^2(s)}
\in {\rm SU(2)}/{\rm U(1)}_3$, 
\begin{eqnarray}
&M_\mu (s)\!\! & \equiv 
\exp\{i\tau ^1\theta _\mu ^1(s)+\tau ^2\theta _\mu ^2(s)\} 
\nonumber\\ 
&\equiv & \! \! \! \! \! \!  \! \! \! \! 
\pmatrix {
\cos\theta _\mu (s)    & \! -e^{-i\chi _\mu (s)}\sin\theta _\mu (s)   \cr 
e^{i\chi _\mu (s)}\sin\theta _\mu (s) & \!  \cos\theta _\mu (s) 
},
\end{eqnarray}
with $-\pi <\theta _\mu ^3(s),\chi _\mu (s) \le \pi $ and 
$0 \le \theta _\mu (s) \le {\pi  \over 2}$. 
Here, $\cos\theta _\mu(s)$ in the abelian gauge 
is a gauge-invariant quantity which measures the `U(1)-ratio' of 
the link-variable $U_\mu(s)$. For instance, 
$\langle \cos\theta _\mu (s) \rangle=1$ means perfectly abelian system.

In MA gauge, the off-diagonal component of $U_\mu^{\rm MA}(s)$ 
is strongly suppressed as $M_\mu (s) \simeq 1$ 
or $U_\mu^{\rm MA}(s) \simeq u_\mu(s)$. 
Actually, the SU(2) lattice QCD in MA gauge shows 
high `U(1)-ratio' as $\langle \cos\theta _\mu (s) \rangle_{\rm MA} \ge 0.9$ 
even in the strong-coupling region. 
Therefore, QCD in MA gauge becomes similar to the abelian 
gauge theory.
\footnote{
In the UV region, such a criterion would be meaningless because 
$U_\mu (s)\rightarrow 1$ as $\beta \rightarrow \infty $ 
in a suitable gauge.
}

For any operator $\hat O[U_\mu (s)]$, 
{\it abelian projection} is realized as
$\langle \hat O[U_\mu (s)]\rangle
\rightarrow  \langle \hat O[u_\mu (s)]\rangle_{\rm MA}$.
In case of $\langle \hat O[U_\mu (s)]\rangle
\simeq \langle \hat O[u_\mu (s)]\rangle_{\rm MA}$, 
the abelian degrees of freedom is relevant for 
$\hat O[U_\mu (s)]$ in MA gauge, which is 
called as {\it abelian dominance} for $\hat O$. 
For instance, the SU(2) lattice QCD shows 
abelian dominance [10] for the string tension as 
$\langle \sigma [u_\mu (s)] \rangle_{\rm MA} \simeq 
0.92 \cdot \langle \sigma [U_\mu (s)] \rangle$.

In U(1)$_3$ link-variable $u_\mu (s)=\exp\{i\tau ^3\theta _\mu ^3(s)\}$, 
$\theta _\mu ^3(s)\in (-\pi ,\pi ]$  is the abelian gauge field on the lattice, 
and {\it abelian field strength} is defined as 
\begin{equation}
\theta _{\mu \nu }^{\rm FS}(s) \equiv 
{\rm mod}_{2\pi }(\partial \wedge \theta ^3)_{\mu \nu }(s)
\in (-\pi ,\pi ], 
\end{equation}
which is U(1)$_3$-gauge invariant. 
Generally, $\theta _\mu ^3(s)$ satisfies 
$
(\partial \wedge \theta ^3)_{\mu \nu }(s) 
=\theta _{\mu \nu }^{\rm FS}(s)+2\pi n_{\mu \nu }(s). 
$
Here, $n_{\mu \nu }(s) \in {\bf Z}$ 
corresponds to the Dirac string and 
varies by singular U(1)$_3$-gauge transformation.
There appear magnetic-monopole currents 
\begin{equation}
k_\nu (s) \equiv {1 \over 2\pi } \partial_\mu  \tilde \theta _{\mu \nu }^{\rm FS}(s)
=-\partial_\mu  \tilde  n_{\mu \nu }(s) \in {\bf Z}
\end{equation}
and the electric current 
$j_\nu (s) \equiv {1 \over 2\pi } \partial_\mu  \theta _{\mu \nu }^{\rm FS}(s)$.

We show in Fig.1 the monopole current $k_\mu (s)$ 
in the lattice QCD in MA gauge. 
In the deconfinement phase, monopole currents only appear as 
short-range fluctuation. 
In the confinement phase, monopole currents cover the whole lattice 
and form a global structure, which is an evidence of 
{\it monopole condensation} [7,8].

Nonperturbative phenomena like confinement are brought by 
large fluctuation of gauge fields in the strong-coupling region.
In MA gauge, such large fluctuation is concentrated into 
the U(1)$_3$ sector, $u_\mu (s)$. 
In particular, monopoles appear at the ends of the Dirac strings, 
and accompany large fluctuation of $u_\mu (s)$ or $\theta _\mu ^3(s)$. 
Hence, monopole density 
$\rho _M \equiv {1 \over V} \sum_{s,\mu}|k_\mu (s)|$ 
is expected to measure the magnitude of gauge-field fluctuation. 
Here, $|k_\mu (s)|$ and $\rho _M$ in MA gauge 
are SU($N_c$)-gauge invariant, since 
$|k_\mu (s)|$ is U(1)$_3$-gauge invariant and Weyl$_2$-invariant.

The abelian gauge field 
$\theta _\mu ^3(s)$ can be decomposed into the {\it monopole part} 
$\theta _\mu ^{Mo}(s)$ and the {\it photon part} $\theta _\mu ^{Ph}(s)$, 
\begin{eqnarray}
\theta _\nu ^{Mo}(s) & \equiv 2\pi  \sum_{s'}
\langle s |\partial^{-2}|s' \rangle 
\partial_\mu  n_{\mu \nu }(s'), \cr
\theta _\nu ^{Ph}(s) & \equiv \sum_{s'}
\langle s |\partial^{-2}|s' \rangle 
\partial_\mu  \theta _{\mu \nu }^{\rm FS}(s'). 
\end{eqnarray}
In the Landau gauge $\partial_\mu \theta _\mu ^3(s)=0$, 
one finds $\theta _\mu ^3(s)=\theta _\mu ^{Mo}(s)+\theta _\mu ^{Ph}(s)$.
U(1)$_3$ link-variables are defined as 
\ 
$u_\mu ^{Mo,Ph}(s) \equiv \exp\{i\tau _3\theta _\mu ^{Mo,Ph}(s)\}$.

>From $\theta _\mu ^{Mo}(s)$ and $\theta _\mu ^{Ph}(s)$, 
one can derive the field strength and the currents 
in the monopole and photon sectors using Eqs.(4) and (5) [11,13]. 
{\it The monopole sector holds the monopole current only} : 
$k_\mu ^{Mo}(s)\simeq k_\mu (s)$ and $j_\mu ^{Mo}(s)\simeq 0$, 
while 
{\it the photon sector holds the electric current only} : 
$j_\mu ^{Ph}(s)\simeq j_\mu (s)$ and $k_\mu ^{Ph}(s)\simeq 0$.

{\it Monopole projection} is realized as 
$\langle \hat O[U_\mu (s)]\rangle$ 
$\rightarrow  \langle \hat O[u_\mu ^{Mo}(s)]\rangle_{\rm MA}$, 
and {\it monopole dominance} as $\langle \hat O[U_\mu (s)]\rangle 
\simeq  \langle \hat O[u_\mu ^{Mo}(s)]\rangle_{\rm MA}$
is observed for NP-QCD. For instance, 
monopole dominance for the string tension [10] is observed 
in the lattice QCD as 
$\langle \sigma [u_\mu ^{Mo}(s)] \rangle_{\rm MA} \simeq 
0.88 \cdot \langle \sigma [U_\mu (s)] \rangle$.

\section{\bf Monopole Dynamics for Confinement}

In MA gauge, 
QCD-monopoles seem essential degrees of freedom for NP-QCD.
In this section, we investigate monopole dynamics and 
confinement properties 
using the monopole-current action [3] extracted 
from the lattice QCD [16], 
\begin{equation}
Z=\sum_{k_\mu (s)\in {\bf Z}} 
\exp\{-\alpha  \sum_s k^2_\mu (s)\}\delta (\partial_\mu k^\mu (s)), 
\end{equation}
which is defined on lattices with large spacing $a$. 
In the dual Higgs phase, 
nonlocal interactions between the monopole current 
would vanish effectively due to the {\it screening effect} [3].

Here, the monopole current with length $L$ is regarded as 
$L$-step {\it self-avoiding random walk} 
with $2d-1$=7 possible direction in each step.
Hence, the partition function (7) is approximated as 
$Z=\sum_L\rho (L)e^{-\alpha L}$ 
$\simeq \sum_L e^{-(\alpha -\ln 7)L}$, 
where the configuration number of 
monopole loop with length $L$ is estimated as 
$\rho (L) \simeq 7^L$. 
In this system, the {\it Kosterlitz-Thouless-type} 
{\it transition} 
occurs at $\alpha _c=\ln 7$ similarly in vortex dynamics 
in the 2-dimensional superconductor.

We perform direct simulations of partition function (7) on 
lattices. 
Fig.2 shows monopole density 
$\rho _M \equiv {1 \over V} \sum_{s,\mu}|k_\mu (s)|$ 
and the clustering parameter $\eta  \equiv \sum_i L_i^2 /(\sum_i L_i)^2$ 
as functions of self-energy $\alpha $. 
As $\alpha $ increases, $\rho _M$ decreases monotonously, 
and {\it declustering of monopole} 
{\it current} is observed 
around $\alpha _c=1.8 \simeq {\rm ln} 7$.

Now, we study confinement in the monopole-current system 
using the dual field formalism [4,5,17]. 
We introduce the dual gauge field $B_\mu $ 
satisfying $\tilde F_{\mu \nu }=(\partial \wedge B)_{\mu \nu }$. 
In the dual Landau gauge $\partial_\mu B^\mu=0$, 
one finds $\partial^2 B_\mu = k_\mu$. 
Hence, starting from the monopole current configuration 
$k_\mu(x)$, the dual gauge field is derived as 
\begin{equation}
B_\mu (x) = \partial^{-2} k_\mu(x) = -{1 \over 4\pi^2}
\int d^4 y {k_\mu(y) \over (x-y)^2}, 
\end{equation}
which leads $F_{\mu \nu }$ and the Wilson loop. 
The Wilson loop $\langle W \rangle$ shown in Fig.3 obeys the area law. 
We show in Fig.4 the string tension $\sigma a^2$ as the function of $\alpha $. 
Similar behavior is found between $\rho _M$ and $\sigma a^2$, 
which suggests the relevant role of monopoles for confinement.

Thus, the monopole theory (7) seems to have essence 
of NP-QCD in the infrared region. 
In real QCD, however, the QCD-monopole would have 
{\it its intrinsic size} $R \sim$ 0.3fm [3], 
for it is a {\it collective mode} composed by gluons. 
Hence, {\it monopole size effects} should appear 
in the UV region, $a \le R$, 
and the monopole action (7) is modified to be {\it nonlocal}. 
In fact, the monopole size $R$ may provide a 
{\it critical scale} for NP-QCD in term of the dual Higgs theory.

\section{\bf Instantons and QCD-monopoles}

The instanton is another relevant topological object in QCD 
according to $\Pi _{3}({\rm SU}(N_c))$ =$Z_\infty$. 
Recent studies reveal close relation between 
instantons and QCD-monopoles [6,8,13-15,18,19]. 
In Fig.5, we show the lattice QCD result for 
the linear correlation between the total monopole-loop 
length $L$ and $I_Q \equiv 
{1 \over 16\pi ^2} \int d^4x |{\rm tr}(G_{\mu \nu } \tilde G_{\mu \nu })|$, 
which corresponds to the total number of 
instantons and anti-instantons. 
The lattice QCD shows also {\it monopole dominance for instantons} 
[13,14]: 
$\langle I_Q[U_\mu ] \rangle 
\simeq \langle I_Q[M_\mu u_\mu ^{Mo}]\rangle_{\rm MA}$
and 
$\langle I_Q[M_\mu u_\mu ^{Ph}]\rangle_{\rm MA} \simeq 0$
\ 
after several cooling. 
Hence, instantons can be regarded as `seeds' of QCD-monopoles 
[6,8,10,13,14,18,19].

For these correlation, 
off-diagonal elements would be essential. 
On the SU(2) lattice in MA gauge, 
we find relatively large off-diagonal elements 
remaining around monopoles. 
Hence, instantons, which need full SU(2) components, 
appear near monopole world-lines in MA gauge.

The authors would like to thank 
Professors Y.~Nambu and R.~Brout 
for their useful comments and discussions.\\

{\bf REFERENCES}

1. Y.~Nambu, Phys.~Rev.~{\bf D10}~(1974)~4262.

2. G.~'t Hooft, Nucl.~Phys.~{\bf B190}~(1981)~455.

3. Z.~F.~Ezawa and A.~Iwazaki, Phys.~Rev. \\
\noindent

\vspace{-0.4cm}
\hspace{0.45cm}{\bf D25} (1982) 2681; {\bf D26}~(1982)~631.

4. H.~Suganuma, S.~Sasaki and H.~Toki, Nucl. \\
\noindent

\vspace{-0.4cm}
\hspace{0.45cm}Phys. {\bf B435} (1995)~207.

5. H.~Suganuma, S.~Sasaki, H.~Toki and \\
\noindent

\vspace{-0.4cm}
\hspace{0.45cm}H.~Ichie, Prog.~Theor.~Phys.~(Suppl.)~{\bf 120}\\
\noindent

\vspace{-0.4cm}
\hspace{0.45cm}(1995)~57.

6. H.~Suganuma, H.~Ichie, S.~Sasaki and \\
\noindent

\vspace{-0.4cm}
\hspace{0.45cm}H.~Toki, {\it Confinement '95}~(World \\
\noindent

\vspace{-0.4cm}
\hspace{0.45cm}Scientific,1995)~65.

7. F.~Brandstater, U.-J.~Wiese and G.~Schier-\\
\noindent

\vspace{-0.4cm}
\hspace{0.45cm}holz,Phys.~Lett~{\bf B272}~(1991)~319.

8. H.~Suganuma, S.~Sasaki, H.~Ichie, H.~Toki \\
\noindent

\vspace{-0.4cm}
\hspace{0.45cm}and F.~Araki, Int.~Symp. on {\it Frontier '96} \\
\noindent

\vspace{-0.4cm}
\hspace{0.45cm}(World Scientific, 1996) 177.

9. A.~Di~Giacomo, Nucl.Phys.{\bf B}~(Proc.Suppl.) \\
\noindent

\vspace{-0.4cm}
\hspace{0.45cm}{\bf 47} (1996) 136 and references therein.

10. M.~Polikarpov, Nucl.~Phys.~{\bf B}(Proc.Suppl.)\\
\noindent

\vspace{-0.4cm}
\hspace{0.6cm}{\bf 53} (1997) 134 and references therein.

11. O.~Miyamura, Phys.~Lett.~{\bf B353} (1995) 91.

12. R.~Woloshyn, Phys.~Rev.~{\bf D51}~(1995)~6411.

13. H.~Suganuma, A.~Tanaka, S.~Sasaki and \\
\noindent

\vspace{-0.4cm}
\hspace{0.6cm}O.~Miyamura, Nucl.~Phys.~{\bf B}(Proc.Suppl.) \\
\noindent

\vspace{-0.4cm}
\hspace{0.6cm}{\bf 47} (1996) 302.

14. H.~Suganuma, S.~Sasaki, H.~Ichie, F.~Araki \\
\noindent

\vspace{-0.4cm} 
\hspace{0.6cm}and O.~Miyamura, Nucl. Phys. {\bf B}\\
\noindent

\vspace{-0.4cm} 
\hspace{0.6cm}(Proc.Suppl.){\bf 53} (1997) 524.

15. S.~Thurner, H.~Markum and W.~Sakuler,\\
\noindent

\vspace{-0.4cm} 
\hspace{0.45cm}
{\it Confinement '95}~(World Scientific,1995)\\
\noindent

\vspace{-0.4cm} 
\hspace{0.6cm}
77.

16. S.~Kitahara, Y.~Matsubara and T.~Suzuki, \\
\noindent

\vspace{-0.4cm} 
\hspace{0.55cm}
Prog.~Theor.~Phys.~{\bf 93} (1995) 1.

17. H.~Ichie, H.~Suganuma and H.~Toki, Phys. \\
\noindent

\vspace{-0.4cm} 
\hspace{0.55cm}
Rev.{\bf D54}(1996)3382; {\bf D52}(1995)2994. 

18. R.~C.~Brower, K.~N.~Orginos and C-I.~Tan,\\
\noindent

\vspace{-0.4cm} 
\hspace{0.55cm} 
Nucl.~Phys.~{\bf B}(Proc.Suppl.){\bf 53}(1997)488.

19. M.~Fukushima, S.~Sasaki, H.~Suganuma, \\
\noindent

\vspace{-0.4cm} 
\hspace{0.55cm}
A.~Tanaka, H.~Toki and D.~Diakonov, \\
\noindent

\vspace{-0.4cm} 
\hspace{0.55cm}
Phys. Lett. {\bf B399} (1997) 141.\\

{\bf FIGURE CAPTION}\\

Fig.1. Monopole current in MA gauge 
extracted from SU(2) lattice QCD 
with $16^3\times 4$. 
(a) confinement phase ($\beta =2.2$), 
(b) deconfinement phase ($\beta =2.4$).\\

Fig.2. (a) Monopole density $\rho _M$ v.s. self-energy $\alpha $. 
\ (b) The clustering parameter $\eta $ 
\  v.s. $\alpha $. 
For $\alpha <\alpha _c$, almost monopole world-lines combine into 
one large cluster ($\eta \simeq 1$).
For $\alpha >\alpha _c$, only small monopole loops appear 
($\eta \simeq 0$).\\

Fig.3. The Wilson loop $\langle W(I,J) \rangle$ 
v.s. $I \times J$ 
in the monopole theory with $\alpha $=1.7, 1.8, 1.9.\\

Fig.4. The string tension $\sigma a^2$ in the monopole-current theory. 
The dotted line denotes the Creutz ratio in the lattice QCD with 
$\beta $=1.25$\alpha $. \\

Fig.5. Correlation between $I_Q$ and the total 
monopole-loop length $L$ in the SU(2) lattice QCD.
We plot the data at 3 cooling sweep on the $16^3 \times 4$ lattice 
with various $\beta$. \\

\end{document}